\documentstyle[twocolumn,seceq,epsf]{jpsj} %%% Added by Nojiri
\def\runtitle{Self-Consistent Second Order Perturbation Theory
for the 2D Hubbard Model}
\def\runauthor{Hidekazu {\sc Nojiri}}

\title
{
Self-Consistent Second Order Perturbation Theory\\
for the Hubbard Model in Two Dimensions
}

\author
{ 
Hidekazu {\sc Nojiri}\footnote{E-mail: j45747a@nucc.cc.nagoya-u.ac.jp}
}

\inst
{
Department of Physics, Nagoya University, Furo-cho, Chikusa-ku, Nagoya 464-8602
}

\recdate
{
\hspace{2cm}
}

\abst
{
We apply self-consistent second order perturbation theory (SCSOPT)
with respect to the on-site repulsive interaction $U$
to study the Hubbard model in two dimensions.
We investigate single particle properties of the model over the entire
doping range at zero temperature.
It is shown that as doping decreases toward half-filling
$\omega$-mass enhancement factor increases,
while $k$-mass enhancement factor decreases.
The increase in $\omega$-mass enhancement factor is larger than
the decrease in $k$-mass enhancement factor, so that 
total-mass is larger than that in the non-interacting case.
When particle number density per unit cell $n$ is given by
$0.64<n<1.0$ interaction enhances anisotropy of the Fermi surface,
whereas at lower densities $n<0.64$ interaction suppresses anisotropy of it.
Due to the decrease in $k$-mass enhancement factor the density of states (DOS)
at the Fermi level is suppressed.
It is possible to understand the results within the framework
of the weak coupling Fermi liquid theory.
}

\kword
{
Hubbard model, two dimensions, Fermi liquid, self consistent, second order perturbation, Fermi surface, $k$-mass, $\omega$-mass
}

\begin{document}
\sloppy
\maketitle

 \section{Introduction}
 Since the discovery of the high-$T_{\rm c}$ superconducting cuprates (HTSC),
~\cite{rf:1}
 two dimensional layered materials have been attracting the attention of many
 researchers.
 Two dimensional Hubbard model is one of the promising model to understand
such materials.
 Investigations of the Hubbard model have a long history and the understanding
 of the one dimensional~\cite{rf:2} and the infinite dimensional cases
 ~\cite{rf:3} is in a higher level.
 In the case of one dimension DOS at the Fermi level becomes zero
with infinitesimal interaction $U$.
 On the other hand in the case of infinite dimensions DOS at the Fermi
level doesn't change at all until metal-insulator transition
 happens due to the locality of the self-energy.
 For the case of two dimensions the feature of the model hasn't been sufficiently
 elucidated.

 We use here the self-consistent second order perturbation theory (SCSOPT)
 with respect to the on-site repulsive interaction $U$
 to study the Hubbard model in two dimensions.
 The SCSOPT satisfies the Luttinger sum rule,
 which is necessary to construct the Fermi liquid theory.~\cite{rf:3.9}
 The SCSOPT was formerly used
 for studying the infinite dimensional case~\cite{rf:4}
 and was known to describe the low energy excitations of the paramagnetic metallic
 phase of the model correctly in the weak coupling case.~\cite{rf:3,rf:4}
 Therefore in the present paper we restrict our attention to the metallic phase
 of the model without magnetic order, which is suggested
 for a non-half-filling case in Quantum Monte Carlo simulations.~\cite
{rf:Hirsch,rf:Furukawa1,rf:Furukawa2}
 On the contrary to infinite dimensional case the self-energy of the two dimensional
 Hubbard model has $k$-dependence.
 The $k$-dependence of the self-energy is considered to play an important role in
 Mott transition in two dimensional systems such as
 ${\rm Ca_{1-x}Sr_xVO_3}$~\cite{rf:5} and HTSC.~\cite{rf:Ino}
 The $k$-dependence effects of the two dimensional Hubbard model were
 examined so far
 by second order perturbation theory (SOPT)
 ~\cite{rf:5.1,rf:Zlatic1,rf:Zlatic2} and
 the fluctuation exchange (FLEX) approximation.~\cite{rf:FLEX1,rf:FLEX2}
 In the case of SOPT it was reported that for sufficiently large $U$
 $n(\mu)$ exceeds unity in spite of $\mu < 0$ around $\mu \simeq 0$, which leads to
 unphysical negative charge susceptibility.~\cite{rf:Zlatic2}
 Then we define $U_1(\mu)$ so that in the region $U > U_1$ SOPT is unphysical.
 In addition SOPT satisfies the Luttinger sum rule only approximately
 for small $U$ except for half-filling. This disagreement between the Fermi surface
 volume and the expectation value of the number operator is also remarkable near
 the half-filling.
 Then we define $U_2$ so that in the region $U > U_2$
 physical features are affected by that disagreement.
 We can use SOPT only in the region $U < min\{U_1, U_2\}$.
 There are not such limitations in SCSOPT.
 Meanwhile the FLEX approximation includes SCSOPT diagrams,
 but it would be meaningful to distinguish the contribution from SCSOPT diagrams
 from the contribution from the other diagrams.

 We adopt the unit system in which $\hbar = 1$ and $k_{\rm B} = 1$.

% Angle resolved photo emission spectra (ARPES).
% One dimensional Hubbard model. Luttinger liquid.
% High-Tc compound.
% In the case of HTSC D $\propto$ $\delta

 \section{Model and Calculations}
 \label{sec:model}
 The Hubbard Hamiltonian is
 \begin{equation}
  \label{eq:2.1}
   H = - t \sum_{n.n., \sigma}
   (c_{i\sigma}^{\dagger}c_{j\sigma} + h.c.)
   + U \sum_{i} n_{i\downarrow}n_{i\uparrow} \ ,
 \end{equation}
% t is the transfer integral of the nearest neighbor sites which reflect
% the locality of the d-electrons
where $t$ is the transfer integral between the nearest neighbor sites and
 $U$ stands for the on-site repulsive interaction.
% of the  Coulomb potential screened by 4s electrons.
% Hereafter we set $t=1$.

The one particle Green's function at zero temperature is defined as~\cite{rf:6.1}
%%% T=0
 \begin{equation}
  \label{eq:2.2}
   G_{{\mib R}\sigma}(t)
   \equiv
   -{\rm i} <T\psi_{\sigma}({\mib R}, t)\psi_{\sigma}^{\dagger}({\mib O}, 0)> \ ,
 \end{equation}
where $\psi_{\sigma}({\mib R}, t)$ denotes a particle field of electrons and
$\left\langle ...\right\rangle $ and $T$ mean thermodynamic average over
grand canonical ensemble and a time ordering operator, respectively.
% \begin{equation}
%  \label{eq:2.3}
%   \psi_{\sigma}({\mib R}, t)
%   \equiv
%   \frac{1}{\sqrt{N_L}}
%   \sum_{{\mib k}}
%   \tilde{c}_{{\mib k}\sigma}
%   e^{i{\mib k} \cdot {\mib R}} 
  % \end{equation}
% \begin{equation}
%  \label{eq:2.4}
%   \tilde{c}_{{\mib k}\sigma}
%   \equiv
%   e^{i\tilde{H}t}c_{{\mib k}\sigma} e^{-i\tilde{H}t}
% \end{equation}
% \begin{equation}
%  \label{eq:2.5}
%   \tilde{H} \equiv
%   H-{\mu}N \ .
% \end{equation}

 The Green's function in $({\mib k}, \omega)$ space can be expressed by
 the self-energy as
 \begin{subequations}
  \begin{equation} 
   \label{eq:2.51.0}
    G_{{\mib k}\sigma}(\omega) =
    \frac{1}
    {\omega - \xi_{\mib k} - \Sigma_{{\mib k}\sigma}(\omega)} \ ,
  \end{equation}
  where
  \begin{equation} 
   \label{eq:2.51.1}
    \xi_{\mib k} = \varepsilon_{\mib k} - \mu
  \end{equation}
  and
  \begin{equation} 
   \label{eq:2.51.2}
    \varepsilon_{\mib k} = -2cos(k_x)-2cos(k_y) \ .
  \end{equation}
 \end{subequations}
 We define here the retarded self-energy and the retarded Green's function as
 \begin{equation} 
  \label{eq:2.52}
   \Sigma_{\mib k}^{(R)}(\omega) \equiv \left\{
				    \begin{array}{ll}
				     \Sigma_{\mib k}(\omega) & \omega > 0 \\
				     \Sigma_{\mib k}^{*}(\omega) & \omega < 0 
				    \end{array}
				   \right. 
 \end{equation}
and
%%% Above line maybe isn't right grammatically ???
%
  \begin{equation} 
   \label{eq:2.53}
    G_{{\mib k}\sigma}^{(R)}(\omega) \equiv
    \frac{1}
    {\omega - \xi_{\mib k} - \Sigma_{{\mib k}\sigma}^{(R)}(\omega)} \ .
  \end{equation}
 The Green's function can be expressed by the spectral function as
 \begin{equation}
  \label{eq:2.6}
   G_{{\mib k}\sigma}(\omega) = 
   \int_{-\infty}^{\infty}{\rm d}E
   \frac{A_{{\mib k} \sigma}(E)}{\omega - E - {\rm i}{\delta}{\rm sign}(E)} \ ,
 \end{equation}
where
 \begin{equation}
  \label{eq:2.7}
  A_{{\mib k} \sigma}(E)
  \equiv
  -\frac{1}{\pi}
  {\rm Im}G_{{\mib k}\sigma}^{(R)}(E) \ .
 \end{equation}
Then it follows 
 \begin{eqnarray}
  \label{eq:2.8}
   G_{{\mib k}\sigma}(t) &=&
   {\rm i}\theta(-t)
   \int_{-\infty}^{0}{\rm d}E
   e^{-{\rm i}Et+{\delta}t}
   A_{{\mib k} \sigma}(E)
   \nonumber\\
  &-&
   {\rm i}\theta(t)\int_{0}^{\infty}{\rm d}E
   e^{-{\rm i}Et-{\delta}t}
   A_{{\mib k} \sigma}(E) 
   \label{eq:2.8.1} \\
  &=&
   {\rm i}\theta(-t)a_{\mib k \sigma}(t)
   -{\rm i}\theta(t)b_{\mib k \sigma}(t) \ .
 \end{eqnarray}
 \begin{subequations}
  Here we define
  \begin{equation}
   \label{eq:2.9}
    a_{\mib k \sigma}(t) \equiv
    \int_{-\infty}^{0}{\rm d}E
    e^{-{\rm i}Et+{\delta}t}
    A_{\mib k \sigma}(E)
  \end{equation}
  and
  \begin{equation} 
   \label{eq:2.10}
    b_{\mib k \sigma}(t) \equiv
    \int_{0}^{\infty}{\rm d}E
    e^{-{\rm i}Et-{\delta}t}
    A_{\mib k \sigma}(E) \ .
  \end{equation}
 \end{subequations}
 Real space Green's functions can be calculated as
 \begin{equation} 
  \label{eq:2.11}
   G_{{\mib R}\sigma}(t)  =
   \frac{1}{N_L}
   \sum_{{\mib k}}
%   \frac{1}{2\pi}
%   \int_{-\infty}^{\infty}d\omega
   G_{{\mib k}\sigma}(t)
   e^{{\rm i} {\mib k} \cdot {\mib R}} \ .
 \end{equation}

\begin{figure}
  %\figureheight{3cm}
%  \epsfxsize=75mm
  \epsfysize=30mm
  $$\epsffile{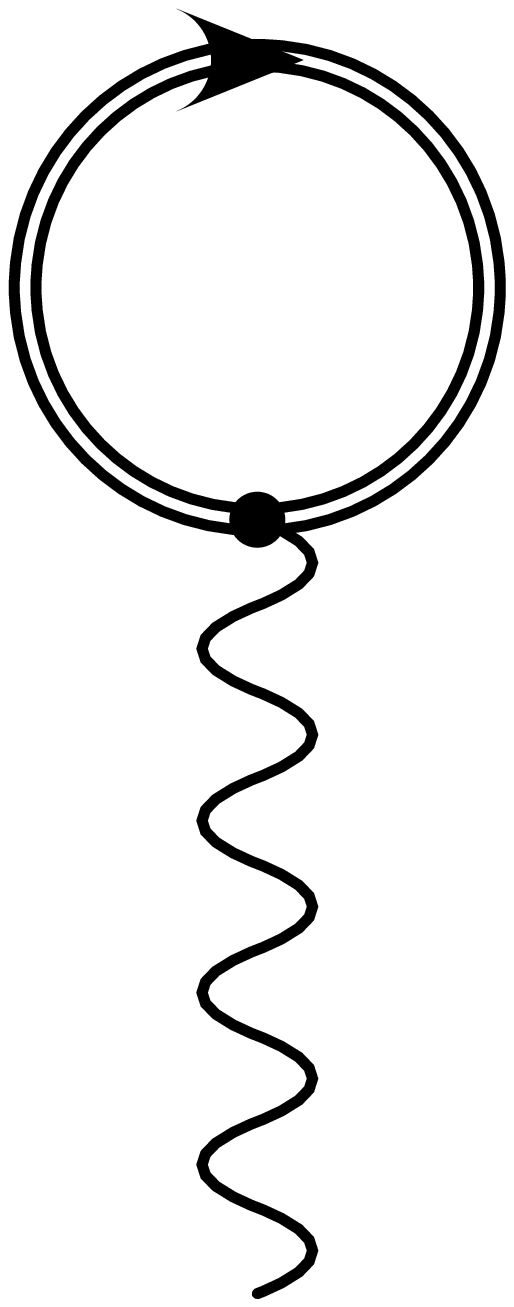}$$
   %\vspace{-30pt}
  \caption{Hartree diagram. The wavy and double solid lines represent the on-site
  repulsive interaction and the exact one particle Green's function, respectively.}
  \label{fig:1}
 \end{figure}

 The contribution of the Hartree diagram (Fig.~\ref{fig:1}) to the self-energy is 
 \begin{equation} 
  \label{eq:2.11.1}
   \Sigma_{{\mib k}\sigma}^{({\rm Hartree})}(\omega) =
   Un_{-\sigma} \ .
 \end{equation}
 As we consider the paramagnetic state in the present paper,
 \begin{equation} 
  \label{eq:2.11.2}
   \Sigma_{{\mib k}\sigma}^{({\rm Hartree})}(\omega) =
   U\frac{n}{2} \ .
 \end{equation}
 This static contribution to the self-energy can be incorporated into
 the chemical potential, and as a result the redefined chemical potential
 takes the value 0 at half-filling.

\begin{figure}
   %\figureheight{4cm}
   %\vspace{-30pt}
%   \epsfxsize=75mm
   \epsfysize=40mm
   $$\epsffile{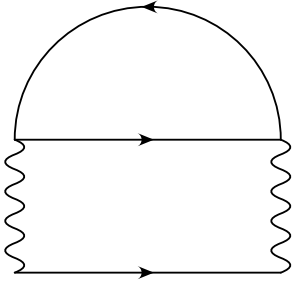}$$
   \caption{Diagrammatic representation of the self-energy in second order in $U$.
   The wavy and the solid lines represent
   on-site repulsive interactions and free Green's functions,
   respectively.}
   \label{fig:3}
  \end{figure}

\begin{figure}
   %\figureheight{4cm}
   %\vspace{-50pt}
%   \epsfxsize=75mm
   \epsfysize=40mm
   $$\epsffile{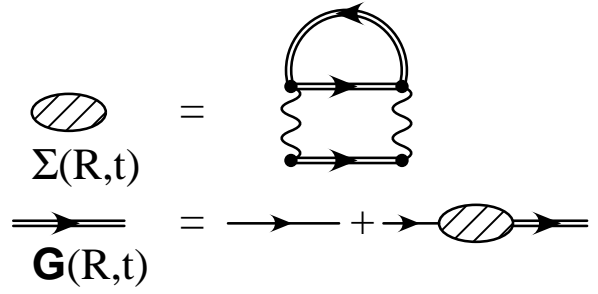}$$
   %\epsfigure{file=fig3.eps,height=4cm}
   %\vspace{-30pt}
   \caption{Diagrammatic representation of SCSOPT.
   The solid, the double solid, and the wavy lines represent
   free Green's functions, renormalized Green's functions, and
   on-site repulsive interactions,
   respectively.}
   \label{fig:2}
  \end{figure}

 The second order contribution (Fig.~\ref{fig:3}) to the
 self-energy is 
\begin{equation} 
 \label{eq:2.12}
  \Sigma_{{\mib R}\sigma}^{(2)}(t) =
  U^{2}
  G_{{\mib R}\sigma}^{(0)}(t)
  G_{\mib R{-\sigma}}^{(0)}(t)
  G_{\mib {-R}{-\sigma}}^{(0)}(-t) \ ,
\end{equation}
where $G_{{\mib R}\sigma}^{(0)}(t)$ is the Green's function of the
non-interacting system.
Similarly the SCSOPT contribution (Fig.~\ref{fig:2}) to the
  self-energy is 
  \begin{equation}
   \label{eq:2.13}
    \Sigma_{{\mib R}\sigma}^{(sc)}(t) =
    U^{2}G_{{\mib R}\sigma}^{(sc)}(t)
    G_{{\mib R}{-\sigma}}^{(sc)}(t)
    G_{\mib {-R}{-\sigma}}^{(sc)}(-t) \ ,
  \end{equation}
  where
  \begin{equation} 
   \label{eq:2.14}
    G_{{\mib k}\sigma}^{(sc)}(\omega) =
    \frac{1}
    {\omega - \xi_{\mib k} - \Sigma_{{\mib k}\sigma}^{(sc)}(\omega)} \ .
  \end{equation}
 In real space the retarded self-energy of SCSOPT is expressed as
 \begin{eqnarray}
  \label{eq:2.16}
   \Sigma_{{\mib R}\sigma}^{(R)(sc)}(t)
   \nonumber &=&
   -{\rm i}U^2\theta(t)\{
   b_{{\mib R}\sigma}(t)
   b_{{\mib R}-\sigma}(t)
   a_{\mib {-R}{-\sigma}}(-t)
   \nonumber\\  &+&
   a_{{\mib R}\sigma}(t)
   a_{{\mib R}{-\sigma}}(t)
   b_{{\mib {-R}}{-\sigma}}(-t) \}
   \nonumber\\  &=&
   U^2\theta(t)\{G_{{\mib R}\sigma}^{(sc)}(t)
   G_{{\mib R}{-\sigma}}^{(sc)}(t)
   G_{{\mib {-R}}{-\sigma}}^{(sc)}(-t)
   \nonumber\\  &+&
   G_{{\mib {-R}}\sigma}^{(sc)*}(-t)
   G_{{\mib {-R}}{-\sigma}}^{(sc)*}(-t)
   G_{{\mib {R}}{-\sigma}}^{(sc)*}(t) \}  \ .\nonumber\\
 \end{eqnarray}
 The real and the imaginary parts of the retarded self-energy satisfy the
 Kramers-Kronig (K.K.) relation,
 \begin{equation}
  \label{eq:2.17}
   {\rm Re}\Sigma_{{\mib k}\sigma}^{(R)}(\omega) =
   \frac{1}{\pi}
   P\int_{-\infty}^{\infty}{\rm d}\omega'
   \frac{{\rm Im}\Sigma_{{\mib k}\sigma}^{(R)}(\omega')}
   {\omega'-\omega} \ .
 \end{equation}

 The algorithm for numerical calculations is as follows.
 \begin{enumerate}
  \renewcommand{\labelenumi}{(\arabic{enumi})}
  \item
       We choose the initial $\Sigma_{{\mib k}\sigma}^{(R)}(\omega)$,
       for example, the self-energy of SOPT.
  \item
       We calculate $G_{{\mib R}\sigma}(t)$ using
       eqs.~(\ref{eq:2.53}),~(\ref{eq:2.7}),~(\ref{eq:2.8.1}) and~(\ref{eq:2.11}).
       At this time, since $A_{\mib k}(\omega)$ is localized
       with respect to energy within
       the extent of the band width, the integration in eq.~(\ref{eq:2.8.1}) can be
       reduced to an appropriate finite energy interval.
  \item
       Using eq.~(\ref{eq:2.16}) we calculate
       $\Sigma_{{\mib R}\sigma}^{(R)}(t)$ which goes to zero more rapidly
       than $\Sigma_{{\mib R}\sigma}(t)$ as $t$ increases.
  \item
       We calculate $\Sigma_{{\mib k}\sigma}^{(R)}(\omega)$ using
       the Fast Fourier Transformation (FFT). At this
       time it is efficient to calculate first
       Im$\Sigma_{{\mib k}\sigma}^{(R)}(\omega)$
       which is localized and calculate
       ${\rm Re}\Sigma_{{\mib k}\sigma}^{(R)}(\omega)$
       using the K.K. relation eq.~(\ref{eq:2.17}).
  \item
       If we iterate this procedure until convergence occurs, we obtain the SCSOPT
       $\Sigma_{{\mib k}\sigma}^{(R)}(\omega)$.
 \end{enumerate}

 Here we present the results for a lattice with $256\times256$ sites
 and the anti-periodic boundary condition.
 The FFT from time to energy variable was performed on 512 points.
 The energy spacing $\delta\omega$ is 0.06 and
 the momentum spacing $\delta{k} \equiv 2\pi/256$ is 0.025.
 All of the numerical results are calculated for zero temperature.
 A typical run to solve the Hubbard model with a fixed chemical potential value
 takes 1 hour on a DEC alpha station 250 4/266.

 Hereafter we drop the spin indices for physical quantities
 as we are in the paramagnetic phase.

%%%
%%% Results
%%%
 \section{Results}
%%%
%%%
  \subsection{Chemical potential shift}
\begin{figure}
   %\figureheight{5cm}
%   \epsfxsize=75mm
   \epsfysize=50mm
   $$\epsffile{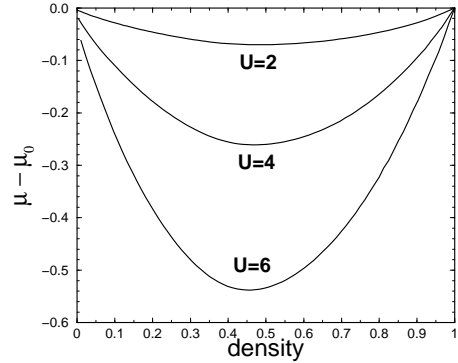}$$
   %\epsfigure{file=fig4.eps,height=5cm}
   \caption{Chemical potential shifts v.s. particle density.
   $\mu$ and $\mu_0$ represent the chemical potential of the interacting and
   the non-interacting system, respectively.}
   \label{fig:5} 
  \end{figure}

  As a function of density the interaction-induced chemical potential shifts
  are shown in Fig.~\ref{fig:5}.
  The chemical potential doesn't shift for half-filling ($n=1$) case
  due to particle-hole symmetry, and has
  a maximal shift $-0.261$ at $n=0.47$ for $U=4$.

  In the weak coupling theory the chemical potential shift
 is related with the $\omega=0$ self-energy
  on the non-interacting Fermi surface as
  ~\cite{rf:7}
  \begin{equation}
   \label{eq:chemical potential shift}
    \delta\mu \simeq
    <{\rm Re}\Sigma_{{\mib k_{F0}}}(0)>_{FS0} \ ,
  \end{equation}
  where ${\mib k_{F0}}$ is the Fermi momentum of the non-interacting system
  and $<...>_{FS0}$ means the average over the non-interacting Fermi surface
  \begin{equation}
   \label{eq:3.4}
   <{\rm Re}\Sigma_{\mib k_{F0}}(0)>_{FS0} \equiv
    \frac{\sum_{\mib k}\delta(\epsilon_{\mib k} - \mu_0)
    {\rm Re}\Sigma_{\mib k_{F0}}(0)}
    {\sum_{{\mib k}}\delta(\epsilon_{\mib k} - \mu_0)} \ ,
  \end{equation}
  where $\mu_0$ is the unperturbed chemical potential.

\begin{figure}
   %\figureheight{5cm}
%   \epsfxsize=75mm
   \epsfysize=50mm
   $$\epsffile{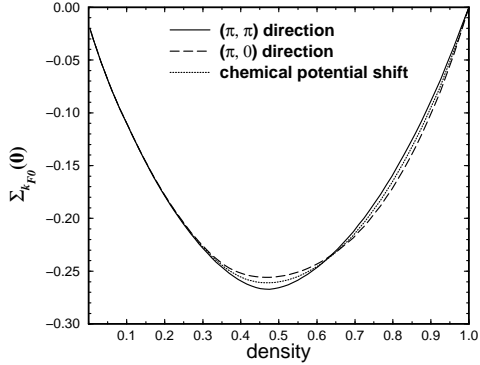}$$
   %\epsfigure{file=fig5.eps,height=5cm}
   %\vspace{-30pt}
   \caption{$\Sigma_{{\mib k_{F0}}}(0)$ of SCSOPT as a function of $n$ 
%   \caption{$\Sigma_{k_{F0}}(0)$ of SCSOPT as a function of $n$ 
   for interaction strength $U=4$.
   The results are shown for $(\pi, 0)$ and $(\pi, \pi)$ directions on the
   non-interacting Fermi surface.
   Chemical potential shift for the same interaction strength
   is also shown for comparison.}
   \label{fig:3.3}
  \end{figure}

  Figure~\ref{fig:3.3} shows the SCSOPT
  ${\rm Re}\Sigma_{\mib k_{F0}}(0)$ as a function of $n$.
  The curve of chemical potential shift is put between
  the curve of Re$\Sigma_{\mib k_{F0}}(0)$ of $(\pi, 0)$ direction and that 
  of $(\pi, \pi)$ direction.
  The variation of ${\rm Re}\Sigma_{\mib k_{F0}}(0)$ on the non-interacting
  Fermi surface is much smaller than ${\rm Re}\Sigma_{\mib k_{F0}}(0)$
  itself.
  Therefore we can conclude that the deviation of $\delta\mu$ from
  eq.~(\ref{eq:chemical potential shift})
  is small compared with $\delta\mu$ itself.
%, and recognize the validity
%  of eq.~(\ref{eq:3.4}).
% which is derived on the weak coupling assumption.
%  reproduces the results.
% Fig.~\ref{fig:3.1}.

%%%
%%%
  \subsection{Fermi surface}
  The excitation energy of the quasi-particle is determined by the
  secular equation 
  \begin{equation}
   \label{eq:3.1}
    E_{\mib k} - \xi_{\mib k} - {\rm Re}\Sigma_{\mib k}(E_{\mib k}) = 0 \ .
  \end{equation}
  The Fermi surface of the interacting system is determined by the condition
$E_{\mib k}=0$, hence
  \begin{equation}
   \label{eq:3.2}
    \xi_{\mib k_F} + {\rm Re}\Sigma_{\mib k_F}(0) = 0 \ ,
  \end{equation}
  where ${\mib k_F}$ is the Fermi momentum of the interacting system.

\begin{figure}
   %\figureheight{5cm}
%   \epsfxsize=75mm
   \epsfysize=50mm
   $$\epsffile{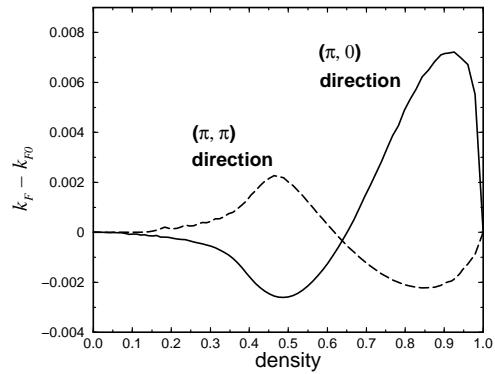}$$
%\epsfigure{file=fig6.eps,height=5cm}
   %\vspace{-30pt}
   \caption{Fermi momentum shifts induced by the interaction 
   over the entire doping range in SCSOPT.
   $k_F$ and $k_{F0}$ denote the Fermi momentum of the interacting and 
   the non-interacting system, respectively.
   The results are shown for $(\pi, 0)$ and $(\pi, \pi)$ directions on the
   Fermi surface.
   Interaction strength $U=4$.}
   \label{fig:3.1}
   %\vspace{-30pt}
  \end{figure}

%  As shown in Fig.~\ref{fig:3.1} for doping range $0.64 < n < 1.0$ the Fermi
%  momentum of the interacting system is larger(smaller) than that of the
%  non-interacting system in $(\pi, 0)((\pi, \pi))$ direction.
%On the other hand 
%  for doping range $n <0.64$ the Fermi momentum of the interacting system is
%  smaller(larger) than that of the non-interacting system in
%  $(\pi, 0)((\pi, \pi))$ direction.
  As shown in Fig.~\ref{fig:3.1} at densities $0.64 < n < 1.0$ the Fermi
  momentum of the interacting system is larger than that of the
  non-interacting system in $(\pi, 0)$ direction, whereas in
  $(\pi, \pi)$ direction it is
  smaller than that of the non-interacting system. On the other hand 
  at densities $n <0.64$ the Fermi momentum of the interacting system
  is smaller than that of the non-interacting system in $(\pi, 0)$ direction,
  whereas in $(\pi, \pi)$ direction it
  is larger than that of the non-interacting system.

\begin{figure}
   %\figureheight{5cm}
%   \epsfxsize=75mm
   \epsfysize=50mm
   $$\epsffile{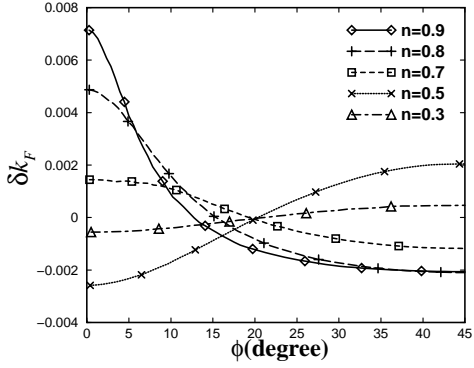}$$
%   \epsfigure{file=fig7.eps,height=5cm}
   %\vspace{-30pt}
   \caption{Fermi surface deformation induced by the interaction
   for various densities.
   The horizontal axis represents the polar angles of positions
   on the non-interacting Fermi surface in units of degree.
   Interaction strength U=4.}
   \label{fig:3.2}
   %\vspace{-30pt}
  \end{figure}

  Figure~\ref{fig:3.2} shows the Fermi surface deformation for various densities.
  $\delta{k_F}$ is defined as 
 \begin{equation}
   \label{eq:3.2999}
   \delta{k_F}({\mib k_{F0}}) {\equiv}
%   ({\mib k_F} - {\mib k_{F0}})_{} \ ,
   ({\mib k_F} - {\mib k_{F0}}) \cdot {{\mib n}_{\mib k_{F0}}} \ ,
 \end{equation}
 where we take the non-interacting Fermi momentum ${\mib k_{F0}}$ so 
  that ${\mib k_{F}} - {\mib k_{F0}}$ is perpendicular to the non-interacting
  Fermi surface, and ${\mib n}_{{\mib k_{F0}}}$ is a unit vector normal
  to the non-interacting
  Fermi surface and always is directed outwards from ${\mib k_{F0}}$.
  $\phi$ represents the polar angles of positions on the
  non-interacting Fermi surface in units of degree,
  which is described as ${\rm Arctan}(k_{F0y}/k_{F0x})$ mathematically.
  At densities 0.9, 0.8 and 0.7, which are larger than 0.64
  $\delta{k_F}$ is monotonously decrease as $\phi$ changes from $0^o$ to $45^o$.
  On the other hand at densities 0.5 and 0.3, which are smaller than 0.64
  $\delta{k_F}$ is monotonously increase in that region of $\phi$.
  Then there is a unique intersection of interacting and non-interacting
  Fermi surfaces in that region of $\phi$ except for the cases of $n=1$
  (due to particle-hole symmetry) and $0.64$.
  That intersection moves from $0^o$ to $22.5^o$ monotonously
  as $n$ changes from 1 to 0.

  Then from Figs.~\ref{fig:3.1} and~\ref{fig:3.2}
%  it is concluded that interaction deforms Fermi surface
  it is concluded that at densities $0.64<n<1.0$
  interaction enhances anisotropy of the Fermi surface,
  whereas at densities $n<0.64$ it suppresses anisotropy.

  Fermi surface deformation is very small and the topology doesn't change.

  In the weak coupling case eq.~(\ref{eq:3.2}) is reduced to~\cite{rf:7} 
 \begin{equation}
   \label{eq:Fermi momentum shift1}
    \delta{k_F}({\mib k_{F0}}) \simeq
    \frac{\delta\mu - {\rm Re}\Sigma_{\mib k_{F0}}(0)}
    {v^0_{\mib k_{F0}}} \ ,
 \end{equation}
 where ${v^0_{\mib k_{F0}}} \equiv |\nabla\varepsilon_{{\mib k_{F0}}}|$
  is the Fermi velocity of the non-interacting system.
  Putting eq.~(\ref{eq:chemical potential shift}) into
  eq.~(\ref{eq:Fermi momentum shift1}) we get 
 \begin{equation}
  \label{eq:Fermi momentum shift2}
  \delta{k_F}({\mib k_{F0}}) \simeq
   \frac{<{\rm Re}\Sigma_{\mib k_{F0}}(0)>_{FS0}-{\rm Re}\Sigma_{\mib k_{F0}}(0)}
    {v^0_{\mib k_{F0}}} \ .
 \end{equation}
%  Halboth {\it et al.}
  Halboth and Metzner calculated the second order coefficient
  of $\delta{k_F}$ in $U$ using eq.~(\ref{eq:Fermi momentum shift2})
  in SOPT.~\cite{rf:7}
  The results of Fig.~\ref{fig:3.2} are almost the same as the results of
  Halboth and Metzner except for the scale of the \mbox{$y$-axis}.
  This means that in SCSOPT eqs.~(\ref{eq:chemical potential shift})
  and~(\ref{eq:Fermi momentum shift1}) are in fact satisfied approximately
  and low energy part
  $<\Sigma_{\mib k_{F0}}^{(2)}(0, \mu_0(n))>_{FS0}
  - \Sigma_{\mib k_{F0}}^{(2)}(0, \mu_0(n))$ used in their work
  is almost the same as
  $<\Sigma_{\mib k_{F0}}^{(sc)}(0, \mu(n))>_{FS0}
  - \Sigma_{\mib k_{F0}}^{(sc)}(0, \mu(n))$
  except for the scale. The difference in the scale originates in the fact
  that in SOPT $\delta{k_F}$ increases in proportion to $U^{2}$, whereas
  in SCSOPT it isn't in proportion to $U^{2}$.

  As shown in Fig.~\ref{fig:3.3}
  at densities $0.64<n<1.0$ Re$\Sigma_{\mib k_{F0}}(0)$ in $(\pi, 0)$
  direction is smaller than that of $(\pi, \pi)$ direction,
  whereas at low densities $n<0.64$ it is larger than that of
  $(\pi, \pi)$ direction.
  These features combined with eq.~(\ref{eq:Fermi momentum shift2}) account
  for the results in Fig.~\ref{fig:3.1}.
%%%%%%%
%%%%%%%
  \subsection{$k$-mass and $\omega$-mass}
  The denominator of the Green's function is expanded around $E_{\mib k}$ as
  \begin{equation}
   \label{eq:3.5}
    \omega - \xi_{\mib k} - {\rm Re}\Sigma_{\mib k}(\omega)
    \simeq
    (1-
    \left.\frac{\partial {\rm Re}\Sigma_{\mib k}(\omega)}{\partial \omega}
    \right|_{\omega = E_{\mib k}}
    )
    (\omega-E_{\mib k}) \ .
  \end{equation}

  Here we define $\omega$-mass and $\omega$-mass enhancement factors as
  \begin{equation}
   \label{eq:3.6}
    \gamma_{\omega} \equiv
    \frac{m_{\omega}}{m_o} \equiv
    1 - \left.\frac{\partial {\rm Re}\Sigma_{\mib k_F}(\omega)}{\partial \omega}
	 \right|_{\omega=0}
     \equiv \frac{1}{Z_{\mib k_F}} \ ,
  \end{equation}
   where $m_o$ is the mass in the case of $U=0$
  \begin{equation}
   \label{eq:3.6.1}
%    v_{\mib k_F}^{0} \equiv
    \frac{k_F}{m_o} \equiv
    |\nabla\xi_{{\mib k_F}}| \ .
  \end{equation}
   We define $k$-mass and $k$-mass enhancement factors as
%   \begin{equation}
%    \label{eq:3.7}
%     \frac{k_F}{m_k} \equiv
%     |{\rm grad}_{\mib k}(\xi_{{\mib k_F}}+{\rm Re}\Sigma_{\mib k_F}(0))| \ ,
%   \end{equation}
   \begin{equation}
    \label{eq:3.7}
     \frac{k_F}{m_k} \equiv
     \left.|{\rm grad}_{\mib k}(\xi_{\mib k}+{\rm Re}\Sigma_{\mib k}(0))|
     \right |_{{\mib k}={\mib k_F}}
   \end{equation}
and
   \begin{equation}
    \label{eq:3.7.1}
     \gamma_{k} \equiv
     \frac{m_k}{m_o} \ .
   \end{equation}
 Dispersion of the quasi-particle energy is expressed in terms of
 mass enhancement factors as
 \begin{equation}
  \label{eq:3.10}
   E_{\mib k} =
   \frac{k_F}
   {m_0\gamma_k\gamma_\omega}
   ({\mib k}-{\mib k_F}) \cdot {{\mib n}_{\mib k_{F}}} \ ,
 \end{equation}
 where we take ${\mib k_F}$ so
  that ${\mib k}-{\mib k_F}$ is perpendicular to the Fermi surface and
  ${\mib n}_{{\mib k_{F}}}$ is a unit vector normal to the Fermi surface
  and always is directed outwards from ${\mib k_{F}}$.

\begin{figure}
   %\figureheight{5cm}
%   \epsfxsize=75mm
   \epsfysize=50mm
   $$\epsffile{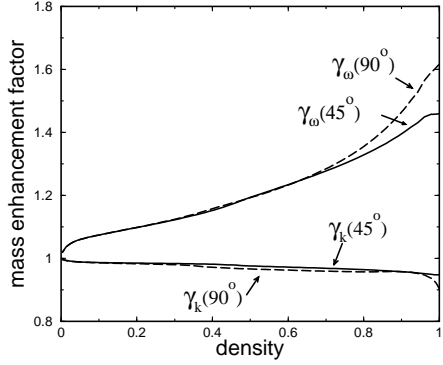}$$
%   \epsfigure{file=fig8.eps,height=5cm}
   \caption{$k$-mass and $\omega$-mass enhancement factors v.s. density of
   particles.
   $\gamma_{\omega}(90^o)$ denotes the $\omega$-mass enhancement factor at the Fermi
   surface in $(0, \pi)$ direction, and $\gamma_{\omega}(45^o)$ denotes the one
   in $(\pi, \pi)$ direction. $\gamma_{k}(90^o)$ and $\gamma_{k}(45^o)$ are
   the $k$-mass enhancement factors defined in the same way.
   Interaction strength $U=4$.}
   \label{fig:7}
  \end{figure}

 Figure~\ref{fig:7} shows that as $n$ increases toward half-filling
 $\omega$-mass enhancement factor increases,
 while $k$-mass enhancement factor decreases
 (Exceptionally $\gamma_{\omega}(90^o)$ increases slightly in the region
 $0.81<n<0.87$).
 The amounts of the increase in $\omega$-mass enhancement factor
 and the decrease in $k$-mass enhancement factor
 are larger in $(0, \pi)$ direction than those in $(\pi, \pi)$ direction.

\begin{figure}
   %\figureheight{5cm}
%   \epsfxsize=75mm
   \epsfysize=50mm
   $$\epsffile{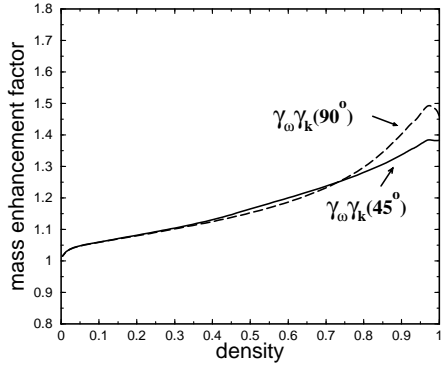}$$
%   \epsfigure{file=fig9.eps,height=5cm}
   \caption{Total mass enhancement factor v.s. density of particles.
   $\gamma_{\omega}\gamma_{k}(90^o)$ denotes the total mass enhancement factor
   at the Fermi surface in $(0, \pi)$ direction,
   and $\gamma_{\omega}\gamma_{k}(45^o)$ denotes that
   in $(\pi, \pi)$ direction. Interaction strength $U=4$.}
   \label{fig:7.1}
  \end{figure}

 From eq.~(\ref{eq:3.10}) total mass enhancement factor is determined
 by $\gamma_\omega\gamma_k$.
 The increase in $\omega$-mass enhancement factor is larger than
 the decrease in $k$-mass enhancement factor, so that total mass is larger than
 that of the non-interacting system. As shown in Fig.~\ref{fig:7.1}
% total mass increases more gradually than $\omega$-mass toward half-filling.
total mass enhancement factor in $(0, \pi)$ direction has a peak at $n=0.97$,
and decreases for $n>0.97$ due to the decrease in $k$-mass enhancement factor.

The decrease in $k$-mass in this weak coupling theory is not sufficient to
explain the experimental data of Mott-Hubbard systems such as 
${\rm Ca_{1-x}Sr_xVO_3}$~\cite{rf:5}, HTSC ${\rm La_{2-x}Sr_xCuO_4}$~\cite{rf:Ino}
and an organic conductor BEDT-TTF salt $\alpha$-${\rm (ET)_2KHg(SCN)_4}$.
~\cite{rf:Sekiyama}

Recently Yoda and Yamada~\cite{rf:Yoda} investigated the effect of the long-range
Coulomb interaction on the $k$-dependence of the self-energy using the perturbation
theory up to the second-order terms. For the half-filling case they found that
the first-order term made a large reduction of the $k$-mass.

On the other hand Ohkawa~\cite{rf:Ohkawa} suggested that in the vicinity of
the Mott transition the formation of light quasi-particle could occur
due to intersite quantum spin fluctuations using 1/d expansion method
within the on-site Coulomb interaction.

The relation between these theories and real systems is not clarified enough now.

%  The origin of the steep decrease of $k$-mass near half-filling
% in the $(0, \pi/2)$ direction is Van-Hove singularity at the point $(0, \pi/2)$.
%  Near the point $(0, \pi/2)$ ${v^0_{\mib k_F}}$ in ~(\ref{eq:3.7}) goes to $0$.
%%%
%%%
%%%
%%%
%%%
  \subsection{Density of States}
  From eq.~(\ref{eq:3.5})
 \begin{equation}
  \label{eq:3.8}
   G_{\mib k}(\omega) \simeq
   \frac{Z_{\mib k_F}}
   {(\omega-E_{\mib k}) - {\rm Im}\Sigma_{\mib k}(\omega)} \ .
 \end{equation}
 If we assume the system can be described by the Fermi liquid theory,
 the coherent part of the spectral function is
 \begin{equation}
  \label{eq:3.9}
   A_{\mib k}(\omega) \simeq
   Z_{\mib k_F}\delta(\omega-E_{\mib k}) \ .
 \end{equation}
 Then the jump in the momentum distribution function across the
 Fermi surface is given by $Z_{\mib k_F}$.
 The density of states of the interacting system is defined as
 \begin{equation}
  \label{eq:3.9.1}
     D(\omega) \equiv
     \frac{1}{N_L}
     \sum_{{\mib k}\sigma}
     A_{\mib k}(\omega) \ .
 \end{equation}
   Putting eqs.~(\ref{eq:3.10}) and~(\ref{eq:3.9}) into eq.~(\ref{eq:3.9.1}),
   we find that density of states at the Fermi level is related with
   the $k$-mass enhancement factor 
   \begin{equation}
    \label{eq:3.11}
     D(0) \simeq
     2\frac{1}{(2\pi)^2}
     \int_{FS}{\rm d}s\frac{\gamma_{k}}{v_{\mib k_F}^{0}} \ ,
   \end{equation}
   where ${v^0_{\mib k_{F}}} \equiv |\nabla\varepsilon_{\mib {k_{F}}}|$
   and $\int_{FS}ds$ denotes the linear integration along the Fermi surface.
   If the system is isotropic in momentum space
   \begin{equation}
  \label{eq:3.12}
   D(0) =
   \gamma_{k}
   D_{U=0}(0) \ .
   \end{equation} 

\begin{figure}
   %\figureheight{5cm}
%   \epsfxsize=75mm
   \epsfysize=50mm
   $$\epsffile{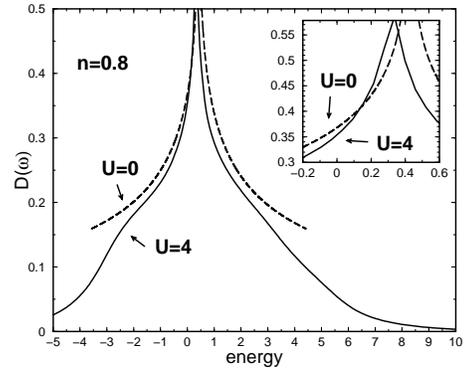}$$
%   \epsfigure{file=fig10.eps,height=5cm,width=8cm}
   %\vspace{-20pt}
   \caption{Renormalized density of states versus energy for $n=0.8$
   of the SCSOPT for interaction strengths $U = 0$ and $4$.
   Inset shows the magnification near the Fermi level.}
   \label{fig:4}
  \end{figure}

 As shown in Fig.~\ref{fig:4} the peak width of DOS
 shrinks as interaction increases, and the peak position of the
 DOS moves toward the Fermi level. These features resemble
 those in the infinite dimensional case.~\cite{rf:4}
 On the other hand DOS at the Fermi level is suppressed with
% the same amount of
 the decrease in the $k$-mass enhancement factor.
 This feature is not found in the infinite
 dimensional case because the self-energy does not have $k$-dependence,
 and consequently DOS at the Fermi level remains the same as that
 of the non-interacting case.~\cite{rf:4}

\begin{figure}
   %\figureheight{5cm}
%   \epsfxsize=75mm
   \epsfysize=50mm
   $$\epsffile{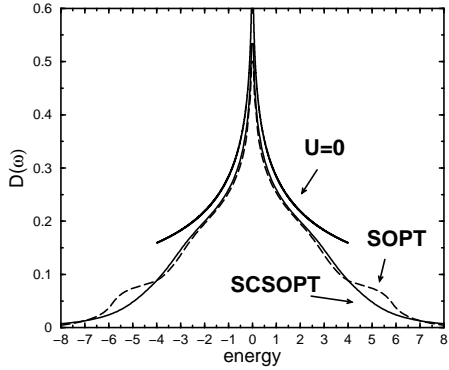}$$
%   \epsfigure{file=fig11.eps,height=5cm,width=8cm}
   %\vspace{-20pt}
   \caption{Renormalized density of states versus energy for $n=1$
   of the SCSOPT and the SOPT (second order perturbation theory)
   for interaction strength $U=4$.
   $U=0$ case is also shown for comparison.}
   \label{fig:4.1}
  \end{figure}

 Figure~\ref{fig:4.1} shows the comparison of the DOS of SCSOPT
 with that of SOPT for $n=1$. In the case of SOPT the weight near the Fermi
 level decreases and two peaks build up at high energies.
 These two peaks are related with the upper and lower
 Hubbard bands.~\cite{rf:7.1,rf:5.1,rf:Zlatic1}
 On the other hand the SCSOPT DOS does not have such two peaks.
 The lack of high-energy resonance peaks in SCSOPT indicates that the SCSOPT is
 a weak coupling theory and is not correct for large $U$.
%Then it is concluded that the SCSOPT is a weak coupling theory.

%%%
%%%
 \subsection{Charge susceptibility}
 If the chemical potential is represented by $\mu_t$, charge susceptibility $\chi_c$
 is defined as ${\rm d}n/{\rm d}{\mu_t}$. Since we define $\mu$ as $\mu_t-Un/2$
 in \S~\ref{sec:model}, charge susceptibility is described as
  \begin{equation}
   \label{eq:3.13}
    \chi_c =
    \frac{{\rm d}n}{{\rm d}\mu}\frac{1}{1+\frac{{\rm d}n}{{\rm d}\mu}
    \frac{U}{2}} \ .
  \end{equation}

  Charge susceptibility for $U=0$ is described as
  \begin{equation}
   \label{eq:3.14}
    \chi_c^0 \equiv
    \frac{{\rm d}n}{{\rm d}\mu_0} =
    2\frac{1}{(2\pi)^2}
    \int_{FS}{\rm d}s\frac{1}{v_{\mib k_F}^0} \ .
  \end{equation}
  For $U=0$, $\chi_c$ diverges logarithmically toward half-filling due to
  the Van-Hove singularity. 

  In the interacting case ${\rm d}n/{\rm d}\mu$ in eq.~(\ref{eq:3.13})
 is described in the Fermi liquid theory as~\cite{rf:Hotta}
  \begin{equation}
   \label{eq:3.15}
    \frac{{\rm d}n}{{\rm d}\mu} =
    \sum_{{\mib k}\sigma}\delta(E_{\mib k}) 
    Z_{\mib k}
    [1 - \frac{\partial {\rm Re}\Sigma_{\mib k}^{(sc)}(0)}{\partial \mu}] \ .
  \end{equation}
  Putting eq.~(\ref{eq:3.10}) into eq.~(\ref{eq:3.15}) we obtain
  \begin{equation}
   \label{eq:3.16}
    \frac{{\rm d}n}{{\rm d}\mu} =
    2\frac{1}{(2\pi)^2}
    \int_{FS}{\rm d}s
    \frac{1}{v_{\mib k_F}^0}
    \gamma_{k}\gamma_{\mu} \ ,
  \end{equation}
  where we define
  \begin{equation}
   \label{eq:3.17}
    \gamma_{\mu} \equiv
    1 - \frac{\partial {\rm Re}\Sigma_{\mib k}(0)}{\partial \mu} \ .
  \end{equation}

\begin{figure}
   %\figureheight{5cm}
%   \epsfxsize=75mm
   \epsfysize=50mm
   $$\epsffile{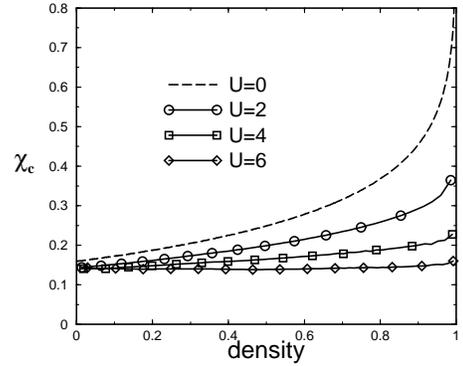}$$
%   \epsfigure{file=fig12.eps,height=5cm}
   \caption{Charge susceptibility $\chi_c$ as a function of the particle density.}
   \label{fig:6}
  \end{figure}

  As shown in Fig.~\ref{fig:6} the charge susceptibility
  $\chi_c$ is suppressed by interaction near half-filling. 
  This suppression is caused
  by the factor $\gamma_{\mu}$ in eq.~(\ref{eq:3.16}), and
  can be thought as the precursor of the Mott insulator charge gap.
  Similar features have been reported in other theories which are in the framework
  of the Fermi liquid theory, such as the theory of infinite dimensional
  Hubbard model~\cite{rf:3} and  second order perturbation theory for
  $\chi_c$ of the two dimensional Hubbard model.~\cite{rf:Hotta}.

  On the other hand our result is inconsistent with Furukawa and Imada's Quantum
  Monte Carlo study~\cite{rf:Furukawa1,rf:Furukawa2} where $\chi_c$
  diverges as $n$ goes to $1$.
%%%%%
%%%%% Conclusion
%%%%%
 \section{Conclusion}
We have applied the self-consistent second order perturbation theory
to study the two dimensional Hubbard model, which is a weak
coupling theory and is expected to describe metallic phase of the model correctly.
The SCSOPT satisfies the Luttinger sum rule exactly,
which is necessary to construct the Fermi liquid theory.

%First we calculate the Fermi surface shift compared
%with the non-interacting system,
First we have calculated the Fermi surface deformation, 
and the result shows that the interaction enhances the anisotropy of the
Fermi surface near half-filling and reduces the anisotropy of it
at lower particle densities.

%It is shown that $\omega$-mass enhancement factor increases toward half filling, 
%on the other hand $k$-mass enhancement factor decreases.
%Because the increase in $\omega$-mass enhancement factor is larger than
%the decrease in $k$-mass enhancement factor, 
%total-mass is larger than that of the non-interacting case.
The peak width of DOS shrinks as interaction increases,
and the peak position of the DOS moves toward the Fermi level
at non-half-filling densities.
These features resemble those in the infinite dimensional case.
On the other hand DOS at the Fermi level is suppressed with
%the same amount of
the decrease in $k$-mass enhancement factor.
The decrease in $k$-mass enhancement factor is small and about
10 percent for $U=4$ at half-filling where it is maximum.
 This feature can not be found in the infinite dimensional case.
%Calculated DOS resembles infinite dimensional case at two points that
%first Kondo resonance peak tend to shrink as interaction increases
%and but on the other hand 

In the weak coupling theory the chemical potential shift induced by the
interaction is related with the average of the $\omega=0$ self-energy over
the non-interacting Fermi surface,
and the Fermi surface deformation is related with the variation of it on
that Fermi surface.
These relations are well satisfied in SCSOPT for the two dimensional Hubbard
model.
%The absolute value of the average
%of the self-energy is at least ten times
%as large as the variation of it.

%The results can be understood by the Green's function formulae
%which assume the weak coupling Fermi liquid state.
%It is possible to understand whole the results within the framework
%of the weak coupling Fermi liquid theory.
%
 \section*{Acknowledgements}
The author is indebted to Professor Y. \={O}no for drawing attention to this
problem and to Professor T. Matsuura and Professor Y. Kuroda
for reading the manuscript and making a number
of helpful suggestions.
He also thanks T. Itakura and N. Nakamura for a lot of discussions.

\end{document}